# SPATIAL–TEMPORAL MESOSCALE MODELING OF RAINFALL INTENSITY USING GAGE AND RADAR DATA

By Montserrat Fuentes,[1] Brian Reich[2] and Gyuwon Lee

*North Carolina State University, North Carolina State University and Kyungpook National University*

Gridded estimated rainfall intensity values at very high spatial and temporal resolution levels are needed as main inputs for weather prediction models to obtain accurate precipitation forecasts, and to verify the performance of precipitation forecast models. These gridded rainfall fields are also the main driver for hydrological models that forecast flash floods, and they are essential for disaster prediction associated with heavy rain. Rainfall information can be obtained from rain gages that provide relatively accurate estimates of the actual rainfall values at point-referenced locations, but they do not characterize well enough the spatial and temporal structure of the rainfall fields. Doppler radar data offer better spatial and temporal coverage, but Doppler radar measures effective radar reflectivity ($Ze$) rather than rainfall rate ($R$). Thus, rainfall estimates from radar data suffer from various uncertainties due to their measuring principle and the conversion from $Ze$ to $R$. We introduce a framework to combine radar reflectivity and gage data, by writing the different sources of rainfall information in terms of an underlying unobservable spatial temporal process with the true rainfall values. We use spatial logistic regression to model the probability of rain for both sources of data in terms of the latent true rainfall process. We characterize the different sources of bias and error in the gage and radar data and we estimate the true rainfall intensity with its posterior predictive distribution, conditioning on the observed data. Our model allows for nonstationary and asymmetry in the spatio-temporal dependency structure of the rainfall process, and allows the temporal evolution of the rainfall process to depend on the motions of rain fields, and the spatial correlation to depend on geographic features. We apply our methods to estimate rainfall intensity every 10 minutes, in a subdomain over South Korea with a spatial resolution of 1 km by 1 km.

Received December 2007; revised February 2008.
[1]Supported in part by NSF Grants DMS-03-53029 and DMS-07-06731.
[2]Supported by NSF Grant DMS-03-54189G.

**1. Introduction.** Precipitation is a key component that links the atmosphere, the ocean and the Earth's surface through complex processes. Thus, accurate knowledge of precipitation levels is a fundamental requirement for improving the prediction of weather systems and of climate change. In particular, some of the weather forecast models use data assimilation techniques that require accurate precipitation estimates at high spatial and temporal resolutions. Accurate rain maps are also used to verify the performance of precipitation forecast models and can pinpoint the shortcomings of numerical models. From a hydrological point of view, the rainfall is a main driver in predicting stream flow. Accurate estimated spatial and temporal structures of precipitation are the most important component to understand and predict flash flood. Rain gages are widely used to measure rainfall accumulation, but the information they provide is limited by their spatial and temporal resolution. In particular, the deployment of dense rain gage networks to resolve detailed spatial structure of rain fields is costly and their maintenance is time-consuming.

Remote sensing technology has improved significantly over the last few decades and can now provide quantitative information on precipitation. Doppler radars estimate instant rainfall intensity or rate ($R$) at very high spatial and temporal resolution (typically, 1 km by 1 km and in the order of minutes). However, a radar does not measure rainfall rate directly but infers the rain rate from the measured effective radar reflectivity ($Ze$). The conversion from radar reflectivity $Ze$ to rainfall $R$ is usually done with the transformation $Ze = \alpha R^\beta$, where $\alpha = 200$, $\beta = 1.6$ [e.g., Lee and Zawadzki (2005)]. This conversion function is typically derived from the measured drop size distributions (DSDs). The DSDs vary with different microphysical and dynamical processes. There is no unique conversion equation that satisfies all different processes, rather varies from storm to storm and within storm [e.g., Lee and Zawadzki (2005)]. The error associated with the $R$–$Ze$ conversion is inherent. Thus, the main source of error of the radar rainfall estimates are associated with reflectivity measurement errors and $R$–$Ze$ conversion errors. Many studies have attempted to correct for these two types of errors [e.g., Zawadzki (1984), Joss and Waldvogel (1990), Jordan, Seed and Austin (2000), Germann et al. (2006), Lee and Zawadzki (2005), Bellon et al. (2006)].

Even when these conversion errors are corrected based on an understanding of the physical processes and the $Ze$ measurement errors are reduced, the radar data are not directly comparable to rain gage measurements due to the differences in sampling volumes between the two different sensors. A rain gage accumulates rainfall at a point on the ground, while radar samples at a volume of approximately 1 km$^3$ at some height above the ground. Rain gages are also relatively sparse across space (in relation to the spatial resolution of the radar data), thus, it is difficult to infer small scale rainfall spatial patterns from gages. Ideally, one would want to combine both sources of rainfall data to obtain more accurate rainfall estimates. This paper introduces a statistical framework for combining radar reflectivity and gage measurements to obtain estimates of rainfall rate, taking into account the different sources of error and bias in both sources of data. We also introduce a sound statistical framework to estimate the $R$–$Ze$ conversion equation as a spatial function. In some of the previous work [e.g., Chumchean, Seed and Sharma (2004)] the relation between radar and gage data is only modeled when it rains, eliminating the zero-rain observations. In our framework, we use all available data, including the zero-rain events. Furthermore, we use spatial logistic regression to model the probability of zero-rain for gage and radar data as a spatial process.

Statistical models have been developed to estimate the distribution of the length of wet and dry periods of rainfall patterns [Green (1964), Hughes and Guttorp (1994), Sanso and Guenni (2000), Zhang and Switzer (2007)]. In particular, the stochastic model introduced by Zhang and Switzer (2007) describes regional-scale, ground-observed storms by using a Boolean random field of rain patches. However, mesoscale instantaneous rainfall patterns are very difficult to model using statistical or physical models, in part due to lack of real-time weather stations measuring rainfall intensity (the Korean network is quite unique in this regard), and because of the spatial and temporal heterogeneity and the inherent variability of rainfall process at high resolution in space and time. This study is aimed to develop a statistical model that can be implemented in real time for the estimation of rainfall intensity maps at high resolutions (1 km in space and 10 minutes in time). Our model treats the true unobservable rainfall intensity as a latent process, and we make inference about that process using a statistical framework that relates radar and gage data and other weather and geographic covariates to the latent process.

Rainfall intensity changes rapidly across space and time, in particular, depending on the direction and intensity of the winds. Estimating complex spatial temporal dependency structures can be computationally demanding, since it involves fitting models that go beyond the standard assumptions of stationarity and full symmetry of the geostatistical models. The assumption of stationarity and full symmetry for spatial–temporal processes offers a simplified representation of any variance–covariance matrix, and consequently, some remarkable computational benefits. Suppose that $\{Z(\mathbf{s},t):\mathbf{s} \equiv (s_1, s_2, \ldots, s_d)' \in \mathbb{D} \subset \mathbb{R}^d, t \in [0,\infty)\}$ denotes a spatial–temporal process where $\mathbf{s}$ is a spatial location over a fixed domain $\mathbb{D}$, $\mathbb{R}^d$ is a $d$-dimensional Euclidean space and $t$ indicates time. The covariance function is defined as $C(\mathbf{s}_i - \mathbf{s}_j; t_k - t_l | \boldsymbol{\theta}) \equiv \operatorname{cov}_{\boldsymbol{\theta}}\{Z(\mathbf{s}_i, t_k), Z(\mathbf{s}_j, t_l)\}$, where $\mathbf{s}_i = (s_1^i, \ldots, s_d^i)'$ and $C$ is positive-definite for all $\boldsymbol{\theta}$, vector of covariance parameters. Under the assumption of stationarity, we have $C(\mathbf{s}_i - \mathbf{s}_j; t_k - t_l | \boldsymbol{\theta}) \equiv$ $C(\mathbf{h}; u|\boldsymbol{\theta})$, where $\mathbf{h} \equiv (h_1, \ldots, h_d)' = \mathbf{s}_i - \mathbf{s}_j$ and $u = t_k - t_l$. Nonstationary models have been introduced by Sampson and Guttorp (1992), Higdon, Swall and Kern (1999), Nychka, Wikle and Royle (2002) and Fuentes (2002), among others. Another assumption of the commonly used spatial models is full symmetry. Under the assumption of full symmetry [Gneiting (2002)] we have $C(\mathbf{h}; u|\boldsymbol{\theta}) = C(\mathbf{h}; -u|\boldsymbol{\theta}) = C(-\mathbf{h}; u|\boldsymbol{\theta}) = C(-\mathbf{h}; -u|\boldsymbol{\theta})$. In the purely spatial context, this property is also known as axial symmetry [Scaccia and Martin (2005)] or reflection symmetry [Lu and Zimmerman (2005)]. Stein (2005) introduced asymmetric models for spatial–temporal data. Park and Fuentes (2008) have introduced this concept of lack symmetry in a more general context for nonstationary space–time process and they have developed asymmetric space–time models. However, these nonstationary and asymmetric models are computationally demanding. In our framework, we allow enough flexibility in the potentially nonstationary and asymmetric spatial–temporal evolution of the rainfall process using a computationally efficient approach. We introduce a latent displacement vector that explains the echo motion of the rainfall intensity, we use motion field information to explain this shift of the rainfall across space and time.

We apply our methods to estimate rainfall intensity at 4:20 am and 4:30 am on July 12, 2006, for a domain over South Korea with a spatial resolution of 1 km by 1 km. The temporal resolution is limited by the update of the radar data (every 10 minutes). At each given time we have 102 observations from the gage network and 10,000 radar reflectivity data points.

This paper is organized as follows. In Section 2 we describe the data. In Section 3 we introduce the statistical framework to combine radar and gage rainfall data. In Section 4 we apply our model to rainfall data in South Korea. We conclude in Section 5 with a discussion and some final remarks.

## 2. Data.

*Radar data.* The Korean Meteorological Administration (KMA) operates 11 Doppler weather radars (4 C-band and 7 S-band) and collects radar volume scans every ten minutes. In this study we use 10 minutes radar data for July 12, 2006 over a subdomain in South Korea at 4:20 am and 4:30 am. The constant altitude Plan Position Indicators (CAPPI) of equivalent radar reflectivity is constructed from each volume scan at 1.5 km height and a temporal and spatial resolution of 10 minutes and 1 km, respectively, after eliminating ground clutter (GC) and anomalous propagation (AP). The reflectivity composite is produced from individual radar CAPPIs with an array of 901 by 1051 at the spatial and temporal resolutions of 1 km and 10 minutes. The weather radars at short wavelengths such as the C-band ($\sim$5 cm) suffer from serious attenuation of signals due to strong precipitation over a radar or along the path of the radar beams. In addition, radar measurements are limited by the blockage of radar beams by complex terrain over the Korean peninsula. The correction of the attenuation and beam blockage could be obtained with complex algorithms that are very computationally demanding. Here, we use a standard approach and we select the maximum value at the overlapping grids to mitigate the attenuation and beam blockage.

The radar composite in South Korea is re-sampled at the smaller domain of 100 km by 100 km around Seoul, South Korea, with latitude ranging from 36.977 to 37.898 and longitude ranging from 126.173 to 127.33. This subdomain particularly provides a dense network of rain gages (see Figure 1). There are two radar stations within this subdomain, at locations ($37°26'$, $126°57'$) and ($37°27'$, $126°21'$) (see Figure 4). The radar on the left-hand side in Figure 4 is C-band so its measurements are affected by attenuation errors. Some of the radar data on the right upper and lower areas of our subdomain are obtained from additional S-band KMA radars located outside our geographic subdomain and, thus, are not shown in Figure 4.

The radar precipitation map is typically generated by converting radar equivalent reflectivity ($Ze$) into rainfall intensity [$R$, units $=$ (mm/h)] using the relationship $Ze = \alpha R^\beta$, $\alpha$ is generally set to 200 and $\beta$ to 1.6. The conversion can result in some error associated with the variation of microphysical processes. The conversion errors vary in time and space [Lee and Zawadzki (2005), Lee et al. (2007)]. In this paper we work directly with the reflectivity $Ze$ and we estimate the values of $\alpha$ and $\beta$, treating $\alpha$ as a spatial process.

*Station data.* The Korean automatic weather stations (AWS) measure the precipitation accumulation, the east–west component of the wind vector (m/s) ($u$ wind component), the north–south component of the wind vector (m/s) ($v$ wind component), temperature (°C) and relative humidity (%) every minute. The bucket size of rain gage is 0.5 mm and the number of tips is recorded every minute from which the 1-minute rainfall intensity (mm/h) is derived using a standard approach called the Tropical Rainfall Measuring Mission (TRMM)/Gauge Data Software Package (GSP) algorithm [Wang, Fisher and Wolf (2008)]. This algorithm first identifies the rain event by checking the time interval between two tips and fills the gap between two tips by adding a half-tip. Then, the cumulative distribution of the amount of rain is derived as a function of time (using cubic splines) and the 1-minute rainfall intensity is obtained from the slope of this function. To minimize the random measurement noise, a 10-minutes moving average centered at the given time is applied, and then the averaged values at the radar measurement time are re-sampled. Zero values represent no-rain but they can also occur by malfunctioning gages. The obtained values are *point* measurements so representative errors need be characterized. In this study we use the weather

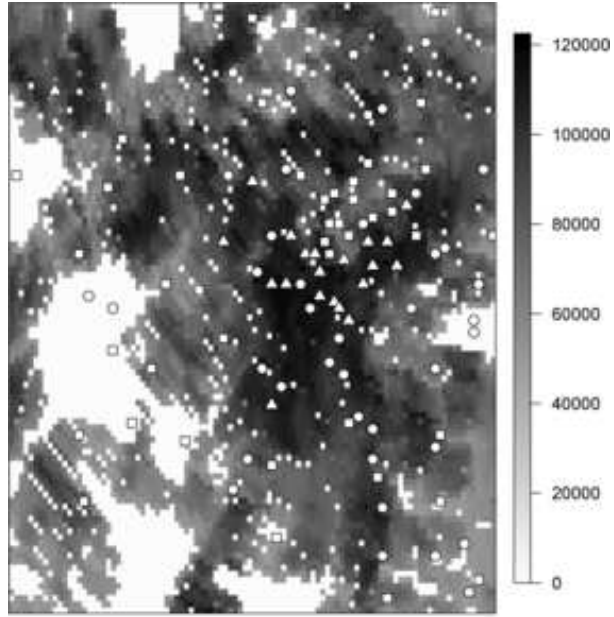

(a) Radar reflectivity and rainfall data at 4:20am.

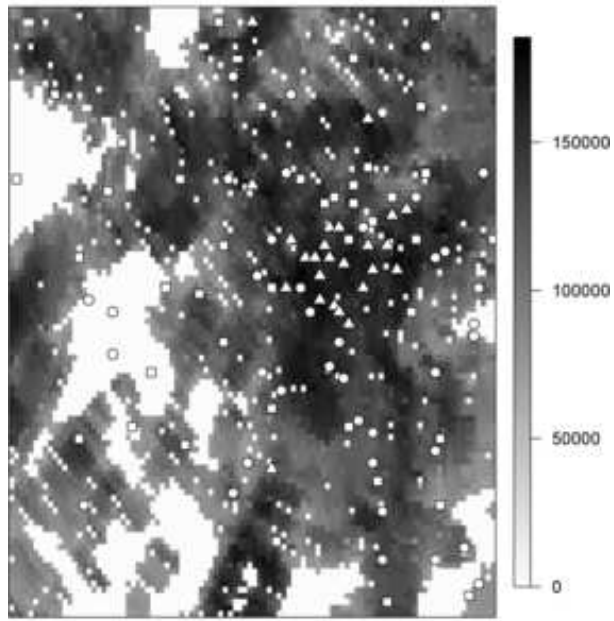

(b) Radar reflectivity and rainfall data at 4:30am.

Fig. 1. *Radar reflectivity and gage data (mm/h) on July 12, 2006 at 4:20 am and 4:30 am. The background color is the reflectivity, the gage values are represented with 3 different symbols: circle = no rain, square = 0–20 mm/h, triangle=more than 20 mm/h.*

stations over a subdomain in South Korea with latitude values ranging from 36.977 to 37.898 and longitude values from 126.173 to 127.33, for July 12, 2006. The total number of weather stations in our subdomain is 102. The climate weather stations in the U.S. record rainfall accumulation every hour and there are only few stations very sparse across space that record real-time meteorological data. Some special stations transmit high temporal resolution in real time. However, all station data used in this study are transmitted every minute in real time. The average spatial resolution is about 13 km and even higher around large cities (few kilometers in some parts of our subdomain). In addition, the weather system in South Korea has complex mesoscale features due to interactions with mountains and oceans. Thus, these station data provide a unique opportunity to explore the small scale structure of rain fields.

*Elevation data.* Geographical data are useful to understand the modulation of precipitation systems by surface conditions, in particular, in Korea where weather systems interact with complex geography. In this study we use Digital Elevation Model (DEM) data. The DEM elevation data (in meters above sea level) have resolution of 1 km and are available from the US Geological Survey. The elevation increases as we move west-to-east in the eastern part of our subdomain. Thus, this phenomenon could lead to more beam blockage for the radar data in these areas, which is compensated by selecting maximum values for the overlapping radar data.

**3. Model description.** Let $\hat{R}(s,t)$ be the continuous spatial temporal process with the observed precipitation values at location $s$ and time $t$ as measured by the rain gages, and $\tilde{R}_i(t)$ the radar reflectivity data for grid cell $i$ and time $t$. $R_i(t)$ represents the continuous component (when it rains) of the underlying true instant rainfall process at time $t$ averaged within grid cell $i$. We model the latent rainfall process $R_i(t)$ in the natural logarithm scale, and we denote $Y_i(t) = \log(R_i(t))$. The logarithm transformation is preferred since in the original scale too much attention is given to strong rainfall rates [Lee and Zawadzki (2005)].

In our framework we have 2 main stages. This framework is fitted using a fully Bayesian approach. In stage 1 we model the different sources of data in terms of an underlying latent rainfall process, and in stage 2 we describe the model for the latent process given some weather covariates. We add a stage 0 to model the weather covariates; this stage is implemented outside the fully Bayesian framework for computational convenience.

*Stage* 0.

*Covariates.* In our framework the weather variables, such as temperature, relative humidity, and wind fields are important covariates to explain the latent rainfall process and its temporal evolution. However, these weather covariates are not observed at all locations of interest for rainfall prediction within our 1 km by 1 km resolution gridded domain. Therefore, we do spatial interpolation using thin plate splines. The number of basis is chosen using generalized cross-validation (GCV) [Craven and Wahba (1979)].

*Stage* 1.

*Gage data model.* We model the gage data in terms of an underlying true rainfall process (averaged within each grid cell) with some error, and we model the probability of zero-rain for the gage data as a spatial process $\pi_g(s,t)$ also in terms of the latent rainfall process. We call this model a zero-inflated log-Gaussian process (LGP). Thus, we have

$$\hat{R}(s,t) = \begin{cases} 0, & \pi_g(s,t), \\ \exp\{Y_{i(s)}(t) + \epsilon_g(s,t)\}, & 1 - \pi_g(s,t), \end{cases}$$

with $\pi_g(s,t)$ the probability of zero rain at location $s$ and time $t$ for the rain gage data. The subindex $i(s)$ for the process $Y$ with the log rainfall values refers to the grid cell containing $s$. The component $\epsilon_g(s,t)$ characterizes the variations in the gage data with respect to the truth at location $s$ and time $t$, due to measurement error and also due in part to the temporal and spatial misalignment (different scales) of the gage data and the truth. The misalignment is a consequence of the temporal averaging of the gage data to obtain $\hat{R}(s,t)$ and the spatial averaging of the truth. This variation is expected to be proportional to the precipitation levels that is why we have multiplicative errors in the original scale.

The logits of the unknown $\pi(s,t)$ probabilities (i.e., the logarithms of the odds) are modeled as a linear function of the rainfall values (log scale) $Y_{i(s)}(t)$, using a spatial logistic regression model,

$$\text{logit}(\pi_g(s,t)) = a_g + b_g Y_{i(s)}(t),$$

where $a_g$ and $b_g$ are unknown coefficients.

In our zero-inflated LGP we model the probability of zero rain in terms of the continuous component of the rainfall, $Y_{i(s)}(t)$, and thus, it differs from most zero-inflated models that treat the probability of zero and the continuous part independently. This is an important feature of our model that allows to smooth some unrealistic nonzero values (due to measurement error) in the middle of a storm.

*Radar reflectivity model.* For the radar reflectivity data we have a similar model, except for the fact that we add multiplicative and additive bias components to the reflectivity–rainfall conversion model,

$$\tilde{R}_i(t) = \begin{cases} 0, & \pi_r(i,t), \\ \exp\{c_1(i) + c_2 Y_i(t) + \epsilon_r(i,t)\}, & 1 - \pi_r(i,t), \end{cases}$$

with $\pi_r(i,t)$ the probability of zero-rain for the radar data at time $t$ and grid cell $i$. The above relationship [when $\tilde{R}_i(t) > 0$] is the reflectivity–rainfall conversion function,

$$\log(\tilde{R}_i(t)) = c_1(i) + c_2 \log(R_i(t)) + \epsilon_r(i,t),$$

with the parameter $c_1$ a spatial function (additive bias term), $c_2$ a multiplicative bias term and $\epsilon_r$ a random error component. We tried to model the parameter $c_2$ as a spatial function too, but that led to a lack of identifiability problem. The reflectivity–rainfall conversion function is modeled in the natural logarithm scale, because the error component is thought to increase with the precipitation values, and in the original scale too much weight is given to high rainfall rates [Lee and Zawadzki (2005), Germann and Zawadzki (2002)]. The logits of the unknown $\pi_r(i,t)$ probabilities are modeled as a linear function of the $Y_i(t)$,

$$\mathrm{logit}(\pi_r(i,t)) = a_r + b_r Y_i(t),$$

where $a_r$ and $b_r$ are unknown coefficients. Modeling the probability of zero rain in terms of the rainfall continuous component becomes very relevant for the radar measurements, since in many occasions we obtain isolated zero-rain events due to radar measurement error in the middle of a storm (see zero-rain radar data in the middle of our geographic domain in Figure 1).

*Stage* 2.

*Latent rainfall process.* The latent rainfall process (log scale) $Y_i(t)$ is modeled as a spatio-temporal Gaussian Markov random field. The process at the first time point $\mathbf{Y}(1) = (Y_1(t), \dots, Y_n(t))'$ has a conditionally autoregressive prior (CAR) prior. The CAR prior for $\mathbf{Y}(1)$ is a multivariate normal with a mean $\mathbf{X}(1)\beta$, for $\mathbf{X}(1) = (X_1(1), \dots, X_n(1))'$, that is a linear function of temperature, relative humidity and elevation at time 1, and with inverse covariance $\tau_Y^2 Q_Y$, where $\tau_Y^2 \sim \mathrm{Gamma}(0.5, 0.005)$ is an unknown precision, and $Q_Y = (I - \rho_Y DW)$, where $\rho_Y \in (0,1)$, $I$ is a diagonal matrix with $I_{ii} = 1$, $W_{ik} = w_{ik} \geq 0$ for $i \neq k$ and $W_{ii} = w_{i+} = \sum_{k \neq i} W_{ik}$, and $D = Diag(1/w_{i+})$ is a diagonal matrix with $D_{ii} = 1/w_{i+}$. An attractive feature of this prior is that the full conditional prior $Y_i(1)|Y_k(1), k \neq i$, is normal with mean $\rho_Y(\sum_{k \neq i} \frac{w_{ik}}{w_{i+}}(Y_k(1) + X_i(1)\beta))$ and variance $1/(\tau^2 w_{i+})$. We assume the weights $w_{jk}$ are known and equal to $w_{jk} = I(j \sim k)$, where $I(j \sim k)$ indicates whether cells $j$ and $k$ are adjacent.

*Temporal evolution of the latent rainfall process.* Successive values of $Y_i(t)$ are modeled using a dynamic linear model [Gelfand, Banerjee and Gamerman (2005)]. The usual dynamic linear model smoothes $Y_j(t)$ toward the cell $j$'s previous value, $Y_j(t-1)$, that is,

$$(1) \qquad Y_i(t) = \rho Y_i(t-1) + X_i(t)\beta + \epsilon_i(t),$$

where $\rho \in (0,1)$ controls the amount of temporal smoothing, $\epsilon_i(t)$ is a multivariate normal vector with CAR prior, and with inverse covariance $\tau^2_{\epsilon(t)} Q_{\epsilon(t)}$ [similar to the CAR model for $\mathbf{Y}(1)$, but with mean zero], $X_i$ is a vector with the following covariates: elevation and the temporal gradients of temperature and relative humidity. However, this is inappropriate for rain data because the storm moves over time. To correctly model the storm dynamics, we introduce a displacement vector $\Delta_i(t)$ that explains the echo motion at the grid point $i$. The model then becomes

$$(2) \qquad Y_i(t) = \rho Y_{i+\Delta_i(t)}(t-1) + X_i(t)\beta + \epsilon_i(t),$$

if the shift $\Delta$ at a given time is not a function of space that would impose one constant translation vector, and it would not allow for rotation. To allow enough flexibility in the temporal evolution of the rainfall process, we model $\Delta$ as a spatial categorical process and we use the wind field information ($u$ and $v$ components) (Figure 2) as relevant covariates to explain the shift. We write $W_{u,i}(t)$ and $W_{v,i}(t)$ to denote the $u$ and $v$ components of the wind at time $t$ and location $i$. $S(t)$ represents a spatial spline basis function at time $t$ [number of basis components is chosen using DIC (deviance information criterion), Speigelhalter et al. (2002)] at time $t$, and $\alpha$, $\beta_1$ and $\beta_2$ are unknown coefficients. The basis function for $S(t)$ are the tensor product of two 1-dimensional cubic B-spline basis functions. This type of basis functions was chosen for computational convenience, while still providing enough smoothing for the term $S(t)$. We used information criteria for the selection of number of basis and we ran sensitivity analysis.

We have

$$\Delta_i(t) = ([\delta_{1,i}(t)], [\delta_{2,i}(t)]),$$

where $[x]$ rounds to the nearest integer to $x$, and

$$\delta_{1,i}(t) = \alpha W_{u,i}(t) + S(t)\beta_1$$

and

$$\delta_{2,i}(t) = \alpha W_{v,i}(t) + S(t)\beta_1.$$

We introduce a spline basis function to characterize the behavior of the storm motion not explained by the wind fields. The wind fields (Figure 2) can be very noisy at this high spatial–temporal resolution, so additional smoothing is needed and that is the role of the spline basis functions.

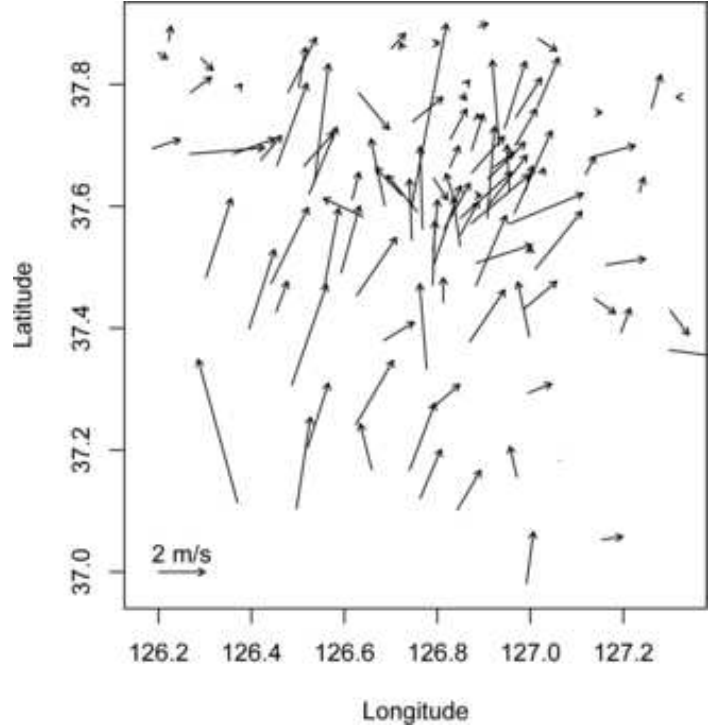

FIG. 2. *Wind fields on July 12, 2006 at 4:20 am. The arrow points in the direction the wind is moving toward and the length of the stem of the arrow is proportional to wind speed.*

We estimate the rainfall intensity, $R_i(t)$, and the probabilities of zero-rain for the gage and radar data, $\pi_g(s,t)$ and $\pi_r(i,t)$, with their predictive posterior distribution summaries. We fit this framework using a fully Bayesian approach. MCMC sampling is carried out using the software $R$ [R Development Core Team (2006)]. Metropolis sampling is used for all parameters. The standard deviations of the Gaussian candidates are tuned to give acceptance probabilities near 0.40. Convergence is monitored by inspecting trace plots of the deviance and several representative model parameters.

3.1. *Covariance function of the rainfall intensity.* It is of interest to calculate the spatial–temporal covariance of the latent rainfall process, as a measure of the dependency structure and variability across time and space of the true underlying rainfall process. The process $\mathbf{Y}(1)$ has inverse covariance $\tau_Y^2 Q_Y$ (defined in the previous section). Conditioning on the shift vector, the covariance of $Y$ (log of the rainfall intensity) is

$\text{cov}(Y_i(t), Y_{i+h}(t+\tau))$

$$= \rho^{2t+\tau} \tau_Y^{-2} Q_Y^{(-1)} \left( i + \sum_{k=2}^{t} \Delta_{i_k^1}(k); i + h \sum_{k'=2}^{t+\tau} \Delta_{i_{k'}^2}(k') \right)$$

$$+ \sum_{j=2}^{t} \rho^{2t-2j+2+\tau} \tau_{\epsilon(j-1)}^{-2} Q_{\epsilon(j-1)}^{(-1)} \left( i + \sum_{k=j}^{t} \Delta_{i_k^1}(k); i + h + \sum_{k'=j}^{t+\tau} \Delta_{i_{k'}^2}(k') \right)$$

$$+ \rho^{\tau} \tau_{\epsilon(t)}^{-2} Q_{\epsilon(t)}^{(-1)} \left( i; i + h + \sum_{k'=t+1}^{t+\tau} \Delta_{i_{k'}^2}(k') \right),$$

where $\rho$ is the smoothing parameter used in equation (2), $Q^{(-1)}(i,j)$ denotes the $(i,j)$ element (row $i$, column $j$) of the inverse matrix of $Q$, and the subindexes $i_k^2$ and $i_k^1$ for $\Delta(k)$ are functions of the location $i$ and values of $\Delta$ at other time points than $k$. When the shift is constant across space, the covariance above becomes a mixture of CAR covariance models. However, by allowing the shift to change across space depending on the wind fields and geographic features, the covariance becomes space-dependent and shows different patterns across space (nonstationarity).

Thus, by using latent processes, we allow the spatial dependency of the log instant rainfall process, $Y_i$, to depend on location, rather than just being modeled as a function of distances between grid cells (stationarity assumption). From the previous expression for the covariance of the process $Y_i$, we see that the temporal evolution of the rainfall process at different locations is a function of the shift vector with the wind fields, allowing for the lack of full symmetry in the space–time covariance.

**4. Application.** The relatively dense spatial coverage of the Korean automatic weather stations with 1-minute weather measurements, combined with the high spatial resolution of the Doppler radar data from KMA, provide a unique opportunity to study the radar reflectivity–rainfall conversion function, and to develop frameworks for spatial temporal mesoscale modeling of rainfall intensity. Lee, Seed and Zawadzki (2007) modeled the spatial/temporal variability of the conversion errors by using reflectivity and rainfall rate measurements at a fixed location. In the work by Lee, Seed and Zawadzki (2007) the conversion errors were not modeled across space and time due to the limited observations. In the work presented here we introduce a statistical model that can capture the complex and heterogeneous spatial temporal structure of mesoscale rainfall intensity combining gage and radar reflectivity data with a resolution of 10 minutes (time) and 1 km ×1 km (space). Figure 1(a) presents the radar reflectivity ($Ze$) and gage rainfall rate ($R$ in mm/h) on July 12, 2006 at 4:20 am. The empirical Pearson correlation between log radar reflectivity and log gage data is 0.5. There is an overall agreement between both sources of information in terms of capturing the large scale structure. However, there are discrepancies at some given locations. Similar empirical correlations are obtained at other time points.

TABLE 1
*We present the DIC value, the posterior expectation of the deviance ($\bar{D}$) and the estimated effective number of parameters $p_D$, for 5 different models.* **Model 1:** *shift and bias constant across space.* **Model 2:** *shift varying spatially and bias constant across space.* **Model 3:** *shift constant across space and bias varying spatially.* **Model 4:** *shift and bias varying spatially (9 spline basis components).* **Model 5:** *shift and bias varying spatially (25 spline basis components)*

| Model | DIC | $\bar{D}$ | $p_D$ |
|---|---|---|---|
| 1 | 18700 | 13092 | 5607 |
| 2 | 18223 | 12535 | 5688 |
| 3 | 19247 | 13377 | 5870 |
| 4 | 17722 | 12020 | 5702 |
| 5 | 18813 | 12436 | 6377 |

Figure 1(b) presents both sources of information on July 12, 2006 at 4:30 am (the next time point). There is a storm moving from the southwestern to the north-eastern part of our domain, and both sources of data seem to capture that phenomenon, showing, however, small scale disagreements in the intensity of the rainfall. We use the statistical framework introduced in this paper to combine radar reflectivity and gage data to estimate rainfall intensity.

To determine and justify the need of the more complex model proposed in this paper, in which we introduce a spatially varying bias function ($c_1$), and a spatially varying shift vector ($\Delta$), we compare 5 different models that assume different spatial structure for the additive bias and the shift vector. In Model 1 we present a simple model, where the shift and bias are constant functions across space. In Model 2 the shift is varying spatially (using the model in Section 3) and the additive bias is constant across space. In Model 3 the shift is constant across space and the bias is varying spatially. In Model 4 the shift and bias are varying spatially this is the general model presented in Section 3, using 9 basis components for the cubic B-spline function (tensor product of two 1-dimensional cubic B-spline basis functions). Model 5 is the same as Model 4 but with a different number of components for the spline function (25 basis components). In Table 1 we present some model comparisons.

We use the DIC [deviance information criterion, Speigelhalter et al. (2002)] to compare model performance. $P_D$ indicates the estimated effective number of parameters. Model 3 has larger $P_D$ than Model 4, even if Model 4 is a more complicated model. The reason for that is that Model 4 has a spatially-varying shift, while Model 3 does not, and adding a small number of shift parameters aligns the two time points in a way that allows for more temporal smoothing and thus fewer effective random effects. Model 4 has the smallest DIC, and we give next a summary of the results for that model.

The prior used for the multiplicative bias (parameter $c_2$ in the reflectivity–rainfall conversion equation) is a Gamma$(1,1)$. The posterior median (95% interval) for this multiplicative bias is 1.05 (1.08, 1.13), which seems to indicate that the recommended 1.6 value for the multiplicative bias (obtained based on regression analysis) might not be always appropriate. In Figure 3 we plot the rainfall gage data versus the reflectivity radar data for our subdomain on July 12,2006 at 4:20 am and 4:30 am, eliminating the no-rain events; we also show the standard conversion curve ($Ze = 200R^{1.6}$). The plot of the spatial posterior median for the spatial bias [parameter $\exp(c_1)$] is presented in Figure 4. The bias is larger in the south-eastern part of our subdomain, where there seems to be more disagreement between radar and gage data. This figure also shows the location of the two radar stations. There is not much relationship between the location of the radar stations and the magnitude of bias for the radar data. It is true that as the radar beams propagate through the *range* of coverage of the Doppler radar, the sampling volume and measurement heights increase. In addition, at larger distances from the station the radar beams can intercept the melting layer of the stratiform rain. Intercepting the melting layer could result in high values of radar reflectivity, mainly due to the increase of the dielectric constant while maintaining the size of the individual precipitation particles. This phenomenon could increase the radar bias at locations that are further away from the station. However, the current storm is mostly a convective system, where the increase of reflectivity in the melting layer is less significant or it is not present. Thus, a larger bias is not expected further away from the radar stations.

The estimated values of $c_2$ were very similar for the other 4 models, which seems to indicate that this parameter is robust to the structure imposed on the additive bias and the shift parameter. The number of basis components for the spline surface is 9 for both the additive bias and also for the shift vector. Models with a different number of spline basis components had higher DIC values (see, e.g., Model 5).

We present in Figure 5 the vector that shows the storm motions ($\Delta_i(t)$). The values presented are the mean of the posterior distribution of the storm direction. We can appreciate in this graph the flexibility offered by our statistical framework to characterize nonstationary and complex patterns in this motion. The estimated displacement vector shows the dominant south–west to north–east direction of the storm, but it also captures some small scales phenomena, such as the west–east shift in the lower left corner of our domain. The average shift is about 11 m/s. The median (95% interval) for the coefficient $\alpha$ that measures the effect of the smoothed wind fields on the shift is 0.39 (0.22, 0.55), which indicates that the smoothed wind plays an important role explaining the storm motion (Figure 5). This might be expected, since storm systems move along the wind fields at steering levels (around 700 to 500 hPa). However, it is also true that the surface winds (Figure 2) are highly affected by local effects such as terrain and local heating. On the other hand, the smoothed version of the wind fields used in this application should attenuate the impact of those local phenomena, and be a better predictor of the shift.

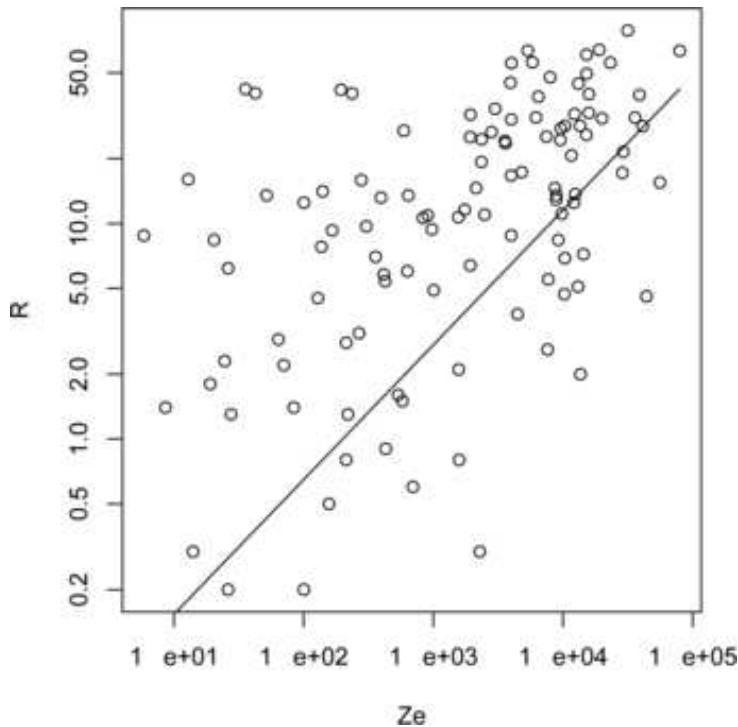

Fig. 3. *Plot of the rainfall data from rain gages, versus radar reflectivity, eliminating the zero-rain events, and using data from July 12, 2006 at 4:20 am and 4:30 am. The solid curve corresponds to the standard transformation $Ze = 200R^{1.6}$. This scatterplot is generated in the natural logarithmic scale for both variables (log of rainfall versus log of reflectivity), presenting the reflectivity radar and gage data on the original scale on the axes.*

It is also of interest to study the probability of rain for both gage and radar data. We used vague normal prior distributions for the logistic parameters, $N(0, 1/0.01)$, where 0.01 is the precision. The median (95% intervals) for the logistic parameters that explain the probability of zero-rain for the radar data are $-2.81$ ($-3.08$, $-2.61$) for the intercept parameter, and $-1.69$ ($-1.83$, $-1.58$) for the slope parameter. For the gage data, we obtain for the intercept $-0.23$ ($-0.68$, $0.40$), and for the slope parameter $-0.17$ ($-0.41$, $0.06$). Both slopes are negative, since higher rain corresponds to less probability of no rain. There are many less gage observations than radar data, so the gage intervals are much wider. In fact, the 95% interval for the slope covers one. Figures 6(a) and (b) show the probabilities of no rain for the radar and gage data. As expected, in the center of our domain near the center of the storm the probability of rain is very high, but there are other areas with high probability in the northwestern part of our domain. The probability of no-rain varies much less across space for the gage data; this may be caused by a higher rate of *erroneous* zero rain values in the gage data than in the radar data. The probability maps illustrate one of the main features of our model, which is the ability to correctly estimate the probability of zero-rain, despite some incorrect zero-rain measurements in radar and gage data in the middle of the storm.

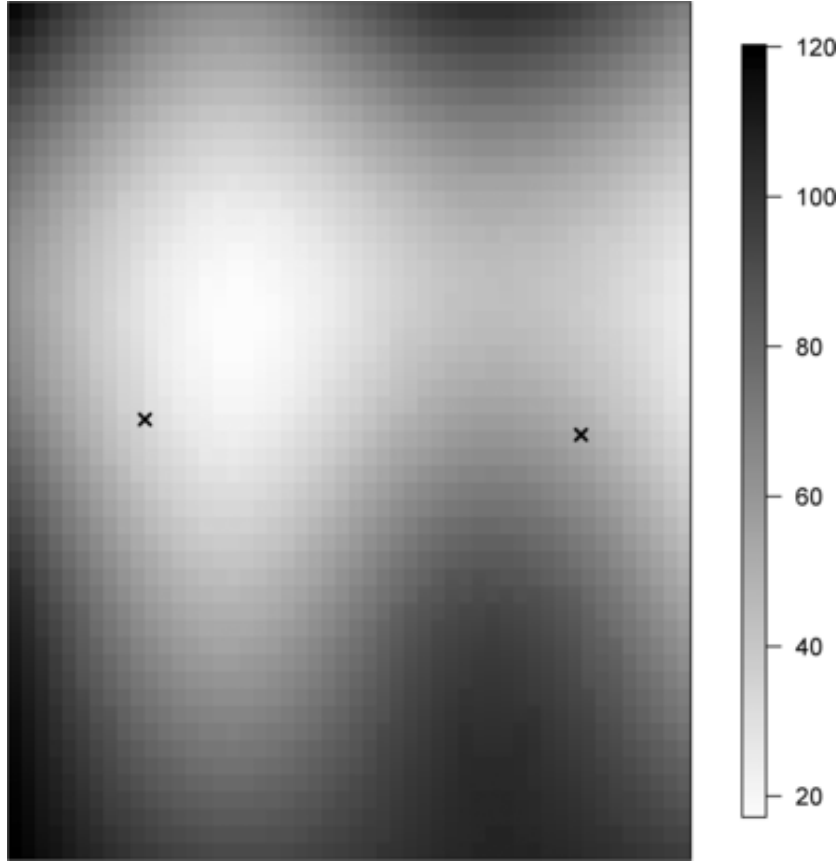

Fig. 4. *Spatial bias surface for the reflectivity radar data at 4.20 am on July 12, 2006 [$\exp(c_1)$ parameter]. The two × marks in the plot indicate the locations of the radar stations within our subdomain.*

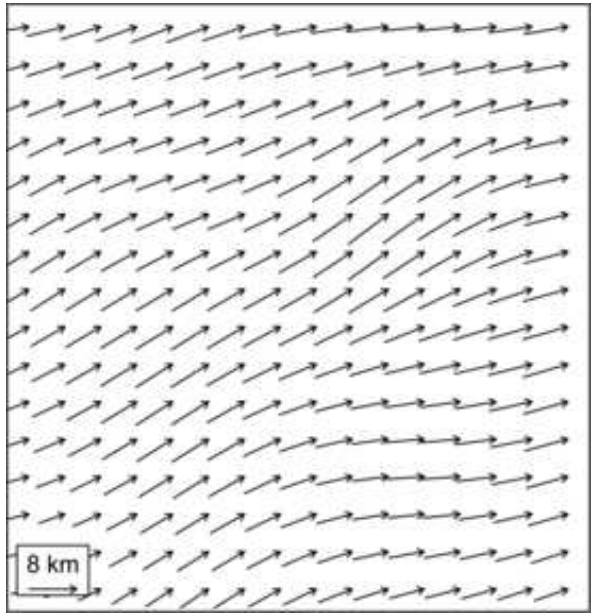

FIG. 5. *The mean of the posterior distribution of the shift vector on July 12, 2006 at 4:30 am.*

We present our rainfall maps in Figures 7(a) and (b). These figures present the mean of the predictive posterior distribution (ppd) of the rainfall intensity, $Z_i(t)$, on July 12, 2006 at 4:20 am and at 4:30 am, respectively. These graphs, overall, present smoother surfaces but with similar rain patterns as the radar reflectivity images (Figure 1), except for the areas with larger bias (see bias function in Figure 4). However, the scale is different, the estimated reflectivity–rainfall function provides the change of scale equation.

We redid the analysis setting the gage values that were thought to be erroneous as missing rather than zero, which is the default value assigned by the recording instrumentation at the gages. An empirical standard approach was used to identify erroneous gage values based on ad-hoc comparison of the gage data with the 9 closest radar pixels. When the number of nonzero rain-value pixels was greater or equal than 3, then the zero-rain gage value was replaced with a *missing* label. The obtained additive bias, shift vectors, and predictive rainfall values [see Figure 7(c) and (d)] were very similar to the ones obtained using the potentially erroneous values. This is due to the fact that in our model we characterize the uncertainty in the gage measurements, and we combine radar and gage data, so clear erroneous values should be

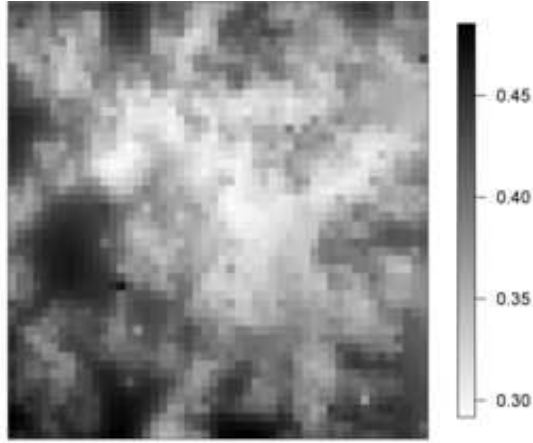

(a) Estimated probability of no-rain for the rain gage data

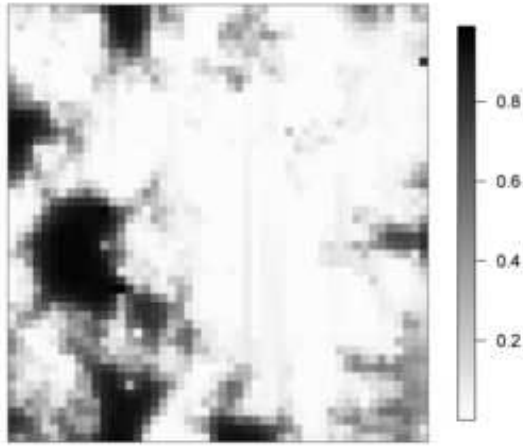

(b) Estimated probability of no-rain for the radar data

FIG. 6. *Graphs (*a*) and (*b*) present the estimated probabilities of no-rain for the rain gage and radar data on July 12, 2006 at 4:20 am.*

detected in our model, which is one of the nice features of our proposed approach.

4.1. *Calibration.* Data from 10% of the gage observations on July 12, 2006 between 4:20 am and 4:30 am are left out for validation. These observations are left out throughout the model fitting. Similarly, 10% of the radar reflectivity data on July 12, 2006 between 4:20 am and 4:30 am are left out for validation. We repeat this calibration analysis 10 times. Thus, we randomly remove 10% of the observations ten separate times and compute

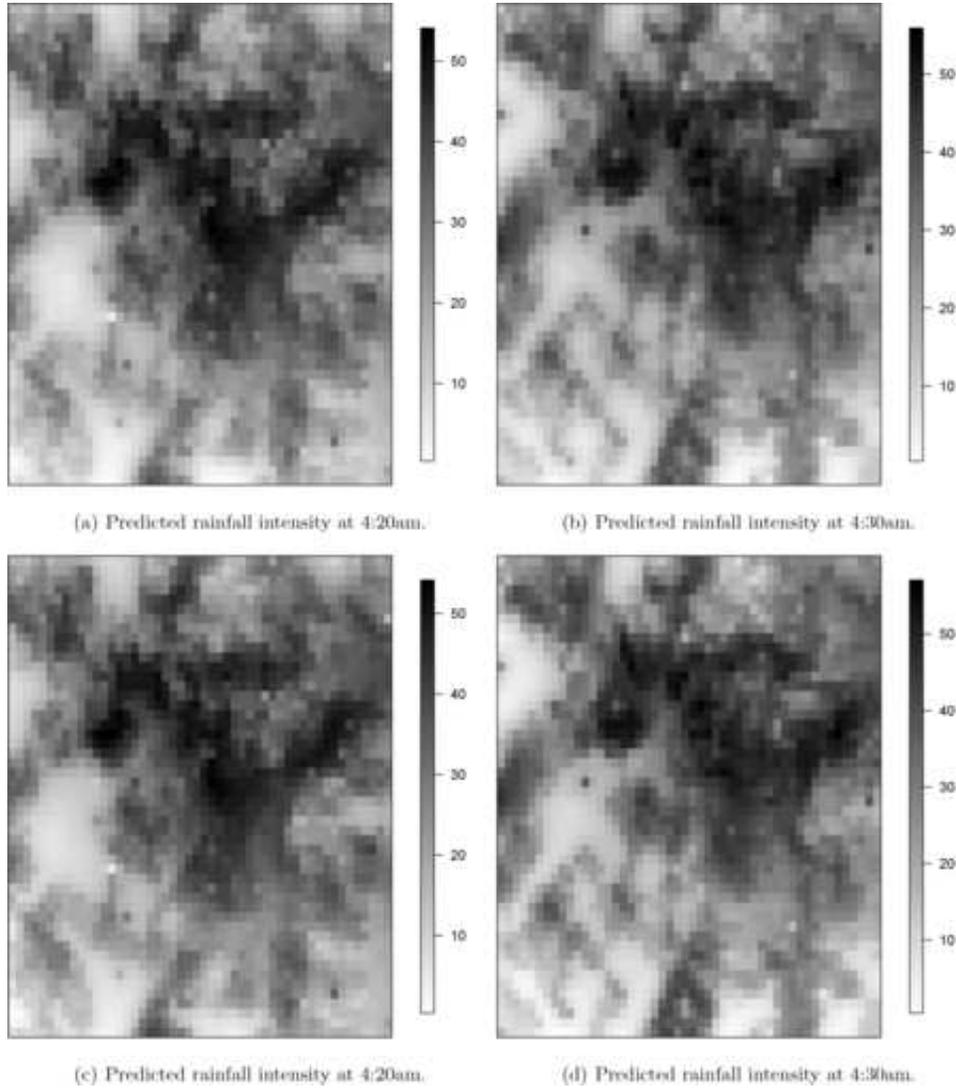

Fig. 7. *Graphs (*a*) and (*b*) present the mean of the predictive posterior distribution of the rainfall intensity (mm/h) on July 12, 2006 at 4:20 am and 4:30 am, respectively. Graphs (*c*) and (*d*) also present the mean of the predictive posterior distribution of the rainfall intensity (mm/h) on July 12, 2006 at 4:20 am and 4:30 am, but, treating what were thought to be erroneous zero-rain gage values as missing rather than zero.*

the coverage probabilities of the 95% prediction intervals for each of the ten analyses.

Figure 8 shows the validation plot for the radar data, for one of the 10 analyses conducted. The predictive reflectivity radar values are 97.4% of the time within the 95% prediction bands. Similar results were obtained for the other 10 analyses. Pooling over all ten analyses, the coverage probabilities are 96% for the radar data and 94% (196/208) for the gage data. We conclude our model is well calibrated.

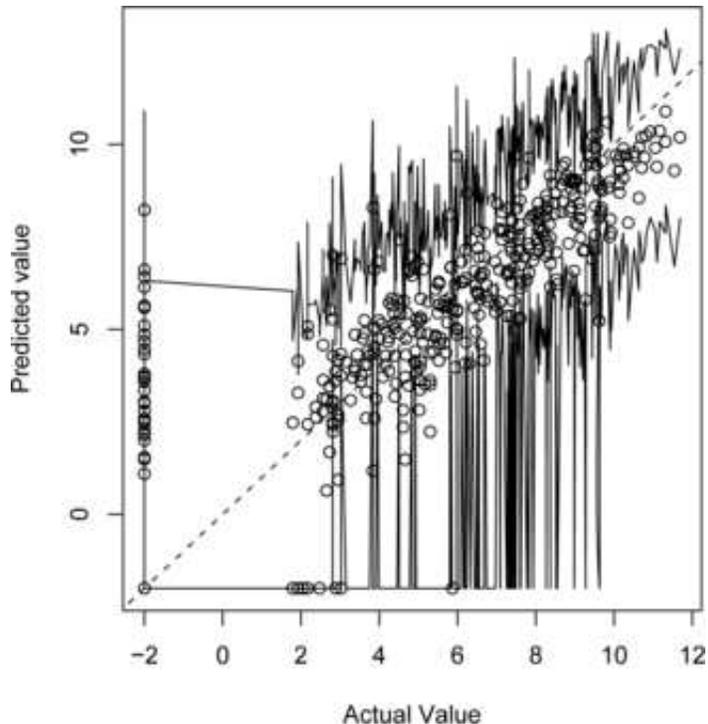

Fig. 8. *Validation plot at randomly selected across space and time hold-out radar data locations on July 12, 2006 between 4:20 and 4:30 am. We hold-out 10% of the radar data. The radar reflectivity data (log scale) versus the predicted values (median of the posterior predictive distribution for log rainfall) are plotted as points, the 95% prediction intervals are plotted as solid-lines. The predictive reflectivity radar values are 97.4% of the time within the 95% prediction bands. The values in the natural logarithm scale are truncated at $-2$. The zero rain values for the observed data are plotted as $-2$.*

**5. Discussion.** In this work we introduce a framework to combine radar reflectivity and gage data to obtain instant rainfall maps. We model the gage and reflectivity data using a zero-inflated log-Gaussian process, with the probability of zero being a function of the latent rainfall process. Most of the previous empirical analyses to study the reflectivity–rainfall conversion equation ignore the zero-rain values. In this work we use the zero-rain values, furthermore, we determine the probability of zero-rain at any given location using a spatial logistic model, taking into account the values of the rainfall process in a neighborhood of the location of interest. The underlying rainfall process has a nonstationary and asymmetric covariance that explains the spatial temporal complex dependency structures.

The dataset used for our analysis seem to indicate that the standard values of the parameters in the reflectivity–rainfall conversion equation might not be appropriate for every storm and geographic domain, and that there is need to allow the additive bias to change across space. The main differences between the approach presented here and previous analyses are the treatment of the zero values, the characterization of the storm motions, and the fact that the additive bias is allowed to change across space. Our model comparisons seem to indicate that these are all important features that lead to a better and more accurate rainfall estimated surfaces.

In this paper we have made an effort to deal with the various problems of the rainfall data sources, but there is always more that could be done. We discuss here some forward-looking issues to illustrate some of the challenges in making the best use of these data. In our modeling framework we have tried to characterize the uncertainty in radar and gage data and in the $R$–$Ze$ conversion equation. However, it would be helpful to conduct different sub-analysis to study and understand better the different sources of error that contribute to the overall uncertainty, in particular, regarding radar mis-calibration and attenuation errors, and errors in the $R$–$Ze$ conversion due to microphysical processes. Radar mis-calibration errors tend to provide a constant bias within the radar coverage range, while errors due to attenuation show high correlation along the path, so it could be possible to separate both sources of error. In addition, $R$–$Ze$ conversion errors should be a function of different microphysical processes. Thus, a new formulation that separates these various errors merits a new exploration. This would provide a better understanding of error sources and a way of mitigating them.

The understanding of error structures and its use in hydrological and meteorological models have enormous potential. Obvious applications are its use in probabilistic verification of the precipitation forecast from numerical models, and of the precipitation estimates from various remote sensing instruments such as the space-based radiometers. From our analysis we can obtain probability maps of exceeding a certain threshold. This information could be used to improve the prediction of the stream flow of extreme hydrological events, and that would aid in the preparation for extreme events. We have shown here the mean of the predictive posterior distributions which shows a central tendency of each predictive field. Numerous predictive fields can be simulated from that distribution and could be used as inputs to generate a hydrological and meteorological ensemble forecast. Since hydrological and meteorological models are nonlinear, the response to the small variation of initial inputs is highly unpredictable. Thus, an ensemble forecast based on some initial predictive fields (as obtained in this paper) that characterize the rainfall variability is essential to characterize extreme events.

**Acknowledgments.** The authors greatly appreciate the Korean Meteorological Administration, in particular, Dr. HyoKyung Kim and Mr. KyungYeup Nam for providing the radar and surface station data. The National Center for Atmospheric Research is sponsored by the National Science Foundation.

M. Fuentes  
B. Reich  
Department of Statistics  
North Carolina State University  
Raleigh, North Carolina 27695  
USA  
E-mail: fuentes@stat.ncsu.edu  
reich@stat.ncsu.edu  

G. Lee  
Department of Astronomy and  
Atmospheric Sciences  
Kyungpook National University  
1370 Sangyeok-dong  
Bukgu, Daegu, 702-701  
Korea (ROK)  
E-mail: gyuwon@knu.ac.kr